\documentclass[preprint,showpacs,preprintnumbers,amsmath,amssymb]{revtex4}
\usepackage{bm}% bold math
\usepackage{epsf}

\everymath{\displaystyle}
\begin{document}           % End of preamble and beginning of text.
%\pagestyle{empty}
%%%%%%%%%%%%%%%%%%%%%%%%%%%%%%%%%%%%%%%%%%%%%%%%%%%%%%%%%%%%%
\title{Quantum-mechanical description of Lense-Thirring effect for relativistic scalar particles}

\author{A.J. Silenko}\email{alsilenko@mail.ru}
\affiliation{Bogoliubov Laboratory of Theoretical Physics, Joint Institute for Nuclear Research,
Dubna, Russia \\ Research Institute for Nuclear Problems,
Belarusian State University, Minsk, Belarus}

\begin{abstract}
Exact expression for the Foldy-Wouthuysen Hamiltonian of scalar particles is used for a quantum-mechanical description of the relativistic Lense-Thirring effect. The exact evolution of the angular momentum operator in the Kerr field approximated by a spatially isotropic metric is found. The quantum-mechanical description of the full Lense-Thirring effect based on the Laplace-Runge-Lenz vector is given in the nonrelativistic and weak-field approximation. Relativistic quantum-mechanical equations for the velocity and acceleration operators are obtained. The equation for the acceleration defines the Coriolis-like and centrifugal-like accelerations and presents the quantum-mechanical description of the frame-dragging effect.
\end{abstract}

%\pacs {03.65.Pm, 04.62.+v, 11.10.Ef, 11.30.-j}
% 03.65.Pm  Relativistic wave equations
% 04.62.+v  Quantum fields in curved spacetime
% 11.10.Ef  Lagrangian and Hamiltonian approach
% 04.20.Jb  Exact solutions
% 11.30.-j  Symmetry and conservation laws

\maketitle

\section{Introduction}

The well-known Lense-Thirring (LT) effect \cite{LT} is a gravitomagnetic effect of frame-dragging predicted by general relativity. It consists in secular precessions of the longitude of the ascending node and the argument of pericenter of a test particle freely orbiting a central spinning mass endowed with angular momentum. This effect also manifests in a precession of the orbit and in a Coriolis-like force acting on the moving particle.

The description of a spinless particle in a Riemannian spacetime of general relativity
is based on the covariant Klein-Gordon-Fock equation \cite{KG} added by an appropriate term describing a nonminimal coupling to the scalar
curvature and conserving the conformal invariance of the equation for a
massless scalar particle \cite{Penrose,CheTagi}. The inclusion of the
Penrose-Chernikov-Tagirov term has been argued for both massive and massless particles
\cite{CheTagi}.

Accioly and Blas
\cite{AccBlas} have brought the initial equation to the Hamiltonian form and have performed the exact Foldy-Wouthuysen (FW) transformation of the Hamiltonian obtained.
They have considered a massive particle in a static isotropic metric. The transformation method used in Ref. \cite{AccBlas} is
inapplicable to massless particles and does not cover
nonstatic spacetimes. As a result, an information about a specific
manifestation of the conformal
invariance in the FW representation has not been obtained.

The generalized method of transformation of the Klein-Gordon-Fock equation to the Hamiltonian form
useful for both massive and massless particles has been developed in Ref. \cite{TMP2008}. Its application in Ref. \cite{Honnefscalar} has allowed to
fulfill the FW transformation and to prove the conformal invariance of the
relativistic FW Hamiltonian for a wide class of
inertial and gravitational fields. General
quantum-mechanical equations of motion have been derived and their classical limit have been obtained.

In the present work, the exact FW Hamiltonian for a scalar particle in the Kerr field approximated by a spatially isotropic metric \cite{Honnefscalar} is used for
a quantum-mechanical description of the relativistic LT effect. We obtain the relativistic equation of motion for the angular momentum operator, perform the quantum-mechanical description of the LT effect in the nonrelativistic and weak-field approximation, and derive quantum-mechanical equations for the velocity and acceleration operators.
The results obtained are compared with the classical description.

\section{Foldy-Wouthuysen Hamiltonian and equations of motion}

The initial covariant Klein-Gordon-Fock
equation with the additional term \cite{Penrose,CheTagi} describes a scalar particle in
a Riemannian spacetime and is given by
\begin{equation}
(\square+m^2-\lambda R)\psi=0,~~~
\square\equiv\frac{1}{\sqrt{-g}}\partial_\mu\sqrt{-g}g^{\mu\nu}\partial_\nu.
\label{eqKG} \end{equation} The Penrose-Chernikov-Tagirov coupling is defined by $\lambda=1/6$. This choice of $\lambda$ has been unambiguously confirmed in Refs. \cite{AccBlas,Honnefscalar}.
Sign of $\lambda$ depends on the definition of $R$. In the present work, the signature is $(+---)$ and the Ricci scalar
curvature is defined by
$R=g^{\mu\nu}R_{\mu\nu}=g^{\mu\nu}R^\alpha_{~\mu\alpha\nu}$, where
$R^\alpha_{~\mu\beta\nu}=\partial_\beta\Gamma^\alpha_{~\mu\nu}-\ldots$
is the Riemann curvature tensor.

The generalized Feshbach-Villars %(GFV)
transformation \cite{TMP2008} and the subsequent nonunitary one allow to represent
Eq. (\ref{eqKG}) in the Hamiltonian form describing both massive and
massless particles \cite{Honnefscalar}:
\begin{equation} \begin{array}{c}  {\cal H}'=\rho_3 \frac{N^2+T'}{2N}+i\rho_2
\frac{-N^2+T'}{2N}-i\Upsilon',\\
%\end{array} \label{eq7i} \end{equation}
%\begin{equation} \begin{array}{c}
T'=\partial_i\frac{G^{ij}}{g^{00}}\partial_j +\frac{m^2-\lambda
R}{g^{00}}+
\frac1f\nabla_i\left(\sqrt{-g}G^{ij}\right)\nabla_j\left(\frac1f\right)\\
+\sqrt{\frac{\sqrt{-g}}{g^{00}}}G^{ij}\nabla_i\nabla_j\left(\frac1f\right)
+\frac{1}{4f^4}\left[\nabla_i(\Gamma^i)\right]^2\\-\frac{1}{2f^2}
\nabla_i\left(\frac{g^{0i}}{g^{00}}\right)\nabla_j\left(\Gamma^j\right)-\frac{g^{0i}}{2g^{00}f^2}
\nabla_i\nabla_j\left(\Gamma^j\right), \\
\Upsilon'\!=\!\frac12\!\left\{\partial_i,\frac{g^{0i}}{g^{00}}\right\},
~~~ G^{ij}\!=\!g^{ij}\!-\!\frac{g^{0i}g^{0j}}{g^{00}},~~~\Gamma^i\!=\!\sqrt{-g}g^{0i},~~~f=\sqrt{g^{00}\sqrt{-g}},
\end{array}\label{eqfr2}\end{equation}
where the nabla operators act only on the operators in brackets
and the primes denote nonunitary transformed operators. Equation
(\ref{eqfr2}) is exact and covers any inertial and
gravitational fields.

The sufficient condition of the exact FW transformation
\cite{TMP2008,JMP,PRA} applied to scalar particles is given by %$[T^\prime,{\cal G}]=0$ or, equivalently,
$\partial_0T^\prime-[T^\prime,\Upsilon']=0$.
%\begin{equation}
%[T^\prime,{\cal G}]=i
%%%$$
%%%\partial_0T^\prime-[T^\prime,\Upsilon']=0. $$
%\label{eFW} \end{equation}
When this condition is satisfied, the exact FW Hamiltonian reads \cite{Honnefscalar}
\begin{equation} {\cal H}_{FW}=\rho_3\sqrt{T^\prime}- i\Upsilon'.
\label{eqFW} \end{equation}
This equation covers \emph{all static spacetimes} ($\Upsilon'=0$) and some important cases of stationary ones.

The metric of the rotating Kerr source has been reduced to the
Arnowitt-Deser-Misner form \cite{ADM} by Hergt and Sch\"afer
\cite{hergt}. This form reproduces the Kerr
solution only approximately.
The form of the metric can be additionally simplified due to
an introduction of spatially isotropic coordinates and
dropping terms violating the isotropy \cite{OSTRONG}:
\begin{equation} \begin{array}{c}
ds^2 = V^2(dx^0)^2 - W^2\delta_{ij}(dx^i - K^idx^0)(dx^j -
K^jdx^0),  ~~~ \bm K=\bm\omega\times\bm r. \end{array}
\label{Ltisotr} \end{equation} The use of the approximate Kerr metric allows to fulfill the
\emph{exact} FW transformation when $V,W$, and $\bm{\omega}$ depend only on the isotropic radial
coordinate $r$. In this approximation, the metric %of the rotating Kerr source takes the form
is defined by
\begin{equation}
\begin{array}{c}   V(r)\!=\!\frac{1-\mu/(2r)}{1+\mu/(2r)}
\! + \!{\cal O}\!\left(\frac{\mu a^2}{r^{3}}\right), ~~~
 W(r)\!=\! \left(1+\frac{\mu}{2r}\right)^{2}\! +\! {\cal O}\!\left(\frac{\mu a^2}{r^{3}}\right), \\
\bm{\omega}(r)= {\frac{2\mu c}{r^3}}\,\bm{a}\left[1 - {\frac
{3\mu}{r}} + {\frac {21\mu^2}{4r^2}}+{\cal
O}\!\left(\frac{a^2}{r^{2}}\right)\right].
\end{array} \label{Okerr}
\end{equation}
Here $\bm{a} = \bm{J}/(Mc),~\mu = GM/c^2$; the total mass $M$ and
the total angular momentum $\bm{J}$ (directed along the $z$ axis)
define the Kerr source uniquely. The leading term in the
expression for $\bm{\omega}(r)=\omega(r)\bm e_z$ corresponds to
the LT approximation.

We can pass on from the Kerr field approximated by Eqs. (\ref{Ltisotr}),(\ref{Okerr}) to a
frame rotating in this field with the angular velocity $\bm o$
after the transformation $dx^i\rightarrow d{X}^i=dx^i\!+\!(\bm
o\times\bm r)dx^0$. The stationary metric of this frame can be
obtained from Eqs. (\ref{Ltisotr}),(\ref{Okerr}) with the replacement
$\bm\omega\rightarrow \bm\Omega\!=\!\bm\omega\!-\!\bm o$. In particular, it covers an
observer on the ground of a rotating source like the Earth or
on a satellite. In this case, $\bm
o=\bm{J}/I$, where $I$ is the moment of inertia. It should be taken into account that frames rotating in the isotropic and Cartesian
coordinates are not equivalent. The exact FW Hamiltonian is given by Eq.
(\ref{eqFW}). When $\lambda=1/6$, the operators $T'$ and $\Upsilon'$ are defined by \cite{Honnefscalar}
\begin{equation}\begin{array}{c} T'=m^2V^2+{\cal F}\bm p^2{\cal
F}-\frac14\nabla{\cal F}\cdot\nabla{\cal F}+\frac16{\cal F}\Delta{\cal
F}+\frac{1}{12}(x^2+y^2)(\Omega'_r)^2,\\
- i\Upsilon'=\bm\Omega\cdot(\bm r\times\bm p),~~~{\cal F}=\frac{V}{W}, \label{eqme}
\end{array} \end{equation}
%Here $\bm L$ is the operator of angular momentum
and derivatives with respect to $r$ are denoted by indexes.$\rightarrow$
In particular, for the LT metric
\begin{equation}\bm\Omega(r)=\frac{2G\bm J}{c^2r^3},~~~V(r)=1-\frac{GM}{c^2r},~~~W(r)=1+\frac{GM}{c^2r}.
\label{LTme}\end{equation}

The quantum-mechanical
equations of motion in the FW representation defining the force, velocity, and acceleration
read $(p_0\equiv {\cal H}_{FW})$
%are defined by the commutators of the FW Hamiltonian with appropriate operators:
\begin{equation}\begin{array}{c}
F^i\!\equiv\!\frac {dp^i}{dt}\!=\frac {\partial p^i}{\partial t}\!+\!\frac{i}{\hbar}\left[{\cal
H}_{FW},p^i\right]=\!\frac12\frac {\partial}{\partial
t}\left\{g^{i\mu},p_\mu\right\}\!+\!\frac{i}{2\hbar}\left[{\cal
H}_{FW},\left\{g^{i\mu},p_\mu\right\}\right],\\
{\cal V}^i\equiv\frac {dx^i}{dt}=\frac{i}{\hbar}\left[{\cal
H}_{FW},x^i\right],~~~ {\cal W}^i=\frac {\partial {\cal
V}^i}{\partial t}\!+\!\frac{i}{\hbar}\left[{\cal H}_{FW},{\cal
V}^i\right].
\end{array}\label{velfactorf}\end{equation}
Any commutation adds the factor $\hbar$ as compared with the
product of operators.

%In the nonrelativistic case, the Wentzel-Kramers-Brillouin (WKB)
%approximation can be used when a de Broglie wavelength is smaller
%than a characteristic size of inhomogeneity region of an external
%field. It has been proved in Ref. \cite{JINRLet1} that satisfying
%this condition allows to use the WKB approximation in the
It has been proved in Ref. \cite{JINRLet1} that satisfying
the condition of the Wentzel-Kramers-Brillouin approximation allows
to use this approximation in the
relativistic case and to obtain a classical limit of the
relativistic quantum mechanics. Determination of the classical
limit reduces to the replacement of operators in the FW
Hamiltonian and quantum-mechanical equations of motion in the FW
representation by respective classical quantities. The
classical limit of the general FW Hamiltonian is given by \cite{Honnefscalar}
\begin{equation}
H = \left(\frac{m^2 - G^{ij}p_ip_j} {g^{00}}\right)^{1/2} -
\frac{g^{0i}p_i}{{g}^{00}}.\label{clCog}
\end{equation}
It coincides with the classical Hamiltonian derived in Ref.
\cite{Cogn}.

The classical limit of Eq. (\ref{velfactorf}) reads
\begin{equation}\begin{array}{c}
{\cal V}^i=\frac{G^{ij}p_j} {\sqrt{g^{00}(m^2 - G^{ij}p_ip_j)}}+
\frac{g^{0i}}{g^{00}},\\
F^i= %%%p_\mu\dot{ g}^{i\mu}+g^{0i}\frac {\partial H}{\partial t}+
p_\mu\frac {\partial g^{i\mu}}{\partial t}+g^{0i}\frac {\partial H}{\partial t}+
g^{ij}\partial_jH+p_\mu {\cal V}^j\partial_j g^{i\mu}.
\end{array}\label{forceclass}
\end{equation}
It %also
coincides with the corresponding classical equations which
follow from Hamiltonian (\ref{clCog}) and the Hamilton equations.
%$$\frac{dH}{dt}=\frac{\partial H}{\partial t},~~~ \frac{dx^{i}}{dt}=-\frac{\partial H}{\partial p_{i}},~~~
%\frac{dp_{i}}{dt}=\frac{\partial H}{\partial x^{i}}.$$
Thus, the quan\-tum-mecha\-ni\-cal and classical equations are in the
best compliance.

For example, the exact metric of a general noninertial frame
characterized by the acceleration $\bm a$ and the rotation $\bm o$
of an observer is defined by $V=1+\bm a\cdot\bm r$, $W=1$, $\bm\Omega=-\bm o$ \cite{HN}.
In this case, the classical limit of the Hamiltonian and
equations of motion is given by \cite{Honnefscalar}
\begin{equation}\begin{array}{c}
H=(1+\bm a\cdot\bm r)\sqrt{m^2+\bm p^2}-\bm o\cdot(\bm r\times\bm p), ~~~
\mbox{\boldmath ${\cal V}$}= (1+\bm a\cdot\bm r)\frac{\bm p}{\sqrt{m^2+\bm p^2}}-\bm o\times\bm r,\\
\mbox{\boldmath ${\cal W}$}=-\bm a(1+\bm a\cdot\bm r)-2\bm o\times\mbox{\boldmath ${\cal V}$}-\bm o\times(\bm o\times\bm r)
+\frac{2\bm a\cdot\mbox{\boldmath ${\cal V}$}+\bm a\cdot(\bm o\times\bm r)}{1+\bm a\cdot\bm r}\left(\mbox{\boldmath ${\cal V}$}+\bm o\times\bm r\right),
\end{array}\label{acclass}
\end{equation} where $\bm p\equiv(-p_1,-p_2,-p_3)$. Leading terms in Eq. (\ref{acclass}) reproduce well-known
classical results \cite{Noninertial}.

\section{Quantum-mechanical description of the Lense-Thirring effect}

The results obtained allow to derive quantum-mechanical equations describing the LT effect. When a metric
depends only on $r$, it is convenient to consider the evolution of the angular momentum operator $\bm l=\bm r\times\bm p$.
Dynamics of this operator in a frame rotating in the Kerr field approximated by a spatially isotropic metric is defined by
\begin{equation} \frac{d\bm l}{dt}=\frac{i}{\hbar}\left[{\cal
H}_{FW},\bm l\right]=\bm\Omega\times\bm l, ~~~ \bm\Omega\!=\!\bm\omega\!-\!\bm o. \label{eqmm}
\end{equation}
Since the operators $\bm\Omega$ and $\bm l$ commute, this equation is exact \emph{for the chosen metric}.

%When we suppose $\bm o=0$, t
The quantity $\bm\omega$ characterizes an evolution of the longitude of the ascending node, $\Upsilon$:
$\Upsilon=\Upsilon_0+\omega t$.
Equations (\ref{Okerr}),(\ref{eqmm}) provide for a relativistic post-Newtonian description of this evolution:
\begin{equation} \omega={\frac{2GJ}{c^2r^3}}\left[1 - {\frac
{3GM}{c^2r}} + {\frac {21G^2M^2}{4c^4r^2}}+{\cal
O}\!\left(\frac{a^2}{r^{2}}\right)\right]. \label{postN}
\end{equation}
This is a part of the LT effect. The longitude of the ascending node can be measured and its measurement is important for astrophysics.

A transition to the classical limit \cite{JINRLet1} and a calculation of the period average in the nonrelativistic and weak-field approximation results in
\begin{equation} \left\langle\frac{1}{r^3}\right\rangle=\frac{1}{b^3(1-e^2)^{3/2}},~~~
\omega={\frac{2GJ}{c^2b^3(1-e^2)^{3/2}}}, \label{nwfa}
\end{equation} where $b$ is the semimajor axis and $e$ is the eccentricity.

The quantum-mechanical description of the full LT effect is based on the Laplace-Runge-Lenz vector.
In this case, we confine ourselves by the nonrelativistic and weak-field approximation. The operator form of the
Laplace-Runge-Lenz vector is given by
\begin{equation} \bm A=\frac12\left(\bm p\times\bm l - \bm l\times\bm p\right)-mk\widehat{\bm r},~~~\widehat{\bm r}=\frac{\bm r}{r}, ~~~ k = GMm. \label{ptN}
\end{equation}
The nonrelativistic FW Hamiltonian for the Kerr field in the LT approximation reads
\begin{equation} {\cal
H}_{FW}=\rho_3\left(mc^2-\frac{k}{r}+\frac{\bm p^2}{2m}\right)+\bm\Omega\cdot\bm l,
\label{nrFWH}\end{equation}
where $\bm\Omega$ is defined by Eq. (\ref{LTme}).
The precession of pericenter of the orbit is defined by the commutator of the operators ${\cal
H}_{FW}$ and $\bm A$:
\begin{equation} \begin{array}{c}
\frac{d\bm A}{\partial t}=\frac{i}{\hbar}[{\cal
H}_{FW},\bm A]=\frac12\left(\bm\Omega\times\bm A-\bm A\times\bm\Omega\right)+\frac{3G}{2c^2}\left\{\frac{\bm J\cdot\bm l}{r^5},\left(\bm r\times\bm l - \bm l\times\bm r\right)\right\}. %\\
%\bm O=\frac{2G}{c^2r^3}\bm J.
\end{array}\label{eqmtA}\end{equation}

The transition to the classical limit and the calculation of the period average leads to the LT equation:
\begin{equation} \begin{array}{c}
\frac{d\bm A}{\partial t}=\bm\Omega_{LT}\times\bm A, ~~~ \bm\Omega_{LT}=\frac{2G}{c^2r^3}\left[\bm J-3(\bm J\cdot\widehat{\bm l})\widehat{\bm l}\right],
\end{array}\label{eqttA}\end{equation}
where $\widehat{\bm l}=\bm l/l$.

The existence of the frame dragging can also be shown. If we hold only main terms in the \emph{relativistic} FW Hamiltonian presented by Eqs. (\ref{eqFW}),(\ref{eqme}), the velocity operator in the field defined by the LT metric (\ref{LTme})
is given by
\begin{equation} \begin{array}{c}
\mbox{\boldmath ${\cal V}$}=\frac{\rho_3}{2}\left\{\frac{c}{\sqrt{m^2c^2V^2+{\cal F}\bm p^2{\cal
F}}},{\cal F}\bm p{\cal
F}\right\}+\bm\Omega\times\bm r.
\end{array}\label{veloper}\end{equation}
In the weak-field approximation, the part of the acceleration operator defined only by the rotation of the source is equal to
\begin{equation} \begin{array}{c}
\mbox{\boldmath ${\cal W}$}=\bm\Omega\times\mbox{\boldmath ${\cal V}$}-\mbox{\boldmath ${\cal V}$}\times\bm\Omega-\bm\Omega\times(\bm\Omega\times\bm r).
\end{array}\label{veper}\end{equation}
This equation defines the Coriolis-like and centrifugal-like accelerations and therefore describes the quantum-mechanical frame-dragging effect.

It is important that the classical limit of all obtained quantum-mechanical equations coincides with the corresponding classical equations.

\section{Conclusions}

The use of the exact FW Hamiltonian for scalar particles
in the frame rotating in the Kerr field approximated by a spatially isotropic metric \cite{Honnefscalar} has allowed us to
fulfill the detailed quantum-mechanical description of the relativistic LT effect. The exact evolution of the angular momentum operator in the Kerr field approximated by a spatially isotropic metric is found. The quantum-mechanical equation defining the precession of pericenter of the orbit (full LT effect) is based on the Laplace-Runge-Lenz vector and derived in the nonrelativistic and weak-field approximation. Relativistic quantum-mechanical equations for the velocity and acceleration operators are obtained. The equation for the acceleration defines the Coriolis-like and centrifugal-like accelerations and presents the quantum-mechanical description of the frame-dragging effect.

The classical limit of the derived quantum-mechanical equations coincides with
corresponding classical ones. This important conclusion confirms the general statement made in Refs. \cite{CheTagi,Tagirov,Honnefscalar} and unambiguously shows a deep connection between the relativistic quantum mechanics of scalar particles in Riemannian spacetimes and the classical general relativity.

\section*{Acknowledgements}

The author is grateful to E. A. Tagirov for his interest in the present study and valuable %comments and
discussions. The work was supported by the
%BRFFR
Belarusian Republican Foundation for Fundamental Research
(Grant No. $\Phi$14D-007).


\begin{thebibliography}{}

\bibitem{LT}
\emph{Thirring H.}
%Uber die Wirkung rotierender ferner Massen in der
%Einsteinschen Gravitationstheorie. [
On the Effect of Rotating Distant
Masses in Einstein's Theory of Gravitation //
Phys. %ikalische
Z. %eitschrift
1918. V 19. P. 33-39 [Gen. Rel. Grav. 1984. V. 16. P. 712-725]; \emph{Thirring H.} Correction to My Paper: ``On the Effect of Rotating
Distant Masses in Einstein's Theory
of Gravitation'' //
Phys. Z. 1921. V. 22. P. 29-30 [Gen. Rel. Grav. 1984. V. 16. P. 725-727];
\emph{Lense J. and Thirring H.}
%Uber den Einfluss der Eigenrotation der
%Zentralkorper auf die Bewegung der Planeten und Monde nach der
%Einsteinschen Gravitationstheorie. [
On the Influence of the Proper Rotation
of Central Bodies on the Motions of Planets and Moons According to Einstein's
Theory of Gravitation //
Phys. Z. 1918. V. 19. P. 156-163
[Gen. Rel. Grav. 1984. V. 16. P. 727-741].
%; English translation
%of both papers as well as their critical analysis is available in: B.
%Mashhoon, F.W. Hehl, and D.S. Theiss, {\it On the gravitational
%effects of rotating masses: The Thirring-Lense papers}, Gen. Relat.
%Grav. {\bf 16} (1984) 711-750.

\bibitem{KG}
\emph{Klein O.}
Quantum Theory and Five-Dimensional Theory of Relativity // Z. Phys. 1926. V. 37. P. 895-906; \emph{Gordon W.} The Compton effect according to Schr\"{o}dinger's theory // Z. Phys. 1926. V. 40. P. 117-133;
\emph{Fock V.} Zur Schr\"{o}dingerschen Wellenmechanik // Z. Phys. 1926. V. 38. P. 242-250.
The equation has been first obtained by E. Schroedinger (unpublished).

\bibitem{Penrose}
\emph{Penrose R.} Conformal Treatment of Infinity // Relativity, Groups and Topology. Edited by C. DeWitt and B. DeWitt. London:
Gordon and Breach, 1964. P. 565-584.

\bibitem{CheTagi}
\emph{Chernikov N.A. and Tagirov E.A.} Quantum theory of scalar field in de Sitter space-time // Ann. Inst. Henri Poincare A. 1968. V. 9. P. 109-141.

\bibitem{AccBlas}
\emph{Accioly A. and Blas H.} Exact Foldy-Wouthuysen transformation for
real spin-0 particle in curved space //
Phys. Rev. D. 2002. V. 66. P. 067501; Conformal coupling and Foldy-Wouthyusen transformation // Mod. Phys. Lett. A. 2003. V. 18. P. 867-873.

\bibitem{TMP2008}
\emph{Silenko A.J.}  Teor. Mat. Fiz. \textbf{156}, 398 (2008) [Hamilton operator and the semiclassical limit for scalar particles in an electromagnetic field // Theor.
Math. Phys. 2008. V. 156. P. 1308-1318].

\bibitem{Honnefscalar}
\emph{Silenko A.J.} Scalar particle in general inertial and gravitational
fields and conformal invariance revisited // Phys. Rev. D. 2013. V. 88. P. 045004; arXiv:1305.6378 [math-ph].

\bibitem{JMP}
\emph{Silenko A.J.} Foldy–Wouthuysen transformation for relativistic particles
in external fields // J. Math. Phys. 2003. V. 44. P. 2952-2966.

\bibitem{PRA}
\emph{Silenko A.J.} Foldy-Wouthyusen transformation and semiclassical limit for relativistic particles
in strong external fields // Phys. Rev. A. 2008. V. 77. P. 012116.

\bibitem{ADM}
\emph{Arnowitt R., Deser S., Misner C.\,W.} The dynamics of general relativity // Gravitation: An
Introduction to Current Research. Edited by L. Witten. New
York: Wiley, 1962. P. 227-265.

\bibitem{hergt}
\emph{Hergt S. and Sch\"afer G.}
Higher-order-in-spin interaction Hamiltonian for binary black holes
from source terms of Kerr geometry in approximate ADM coordinates //
Phys. Rev. D. 2008. V. 77. P. 104001. % (15 pages).

\bibitem{OSTRONG}
\emph{Obukhov Yu.\,N.,  Silenko A.\,J.,  Teryaev O.\,V.} Spin dynamics in gravitational fields of rotating bodies and the equivalence principle // Phys. Rev. D.
2009. V. 80. P. 064044 (2009); Dirac fermions in strong gravitational fields //
Phys. Rev. D. 2011. V. 84. P. 024025.  %Iss. 2. P. .

\bibitem{JINRLet1}
\emph{Silenko A.J.} Pis'ma Zh. Fiz. Elem. Chast. Atom. Yadra
\textbf{10}, 144 (2013) [Classical Limit of Equations of the Relativistic Quantum Mechanics in the Foldy-Wouthuysen Representation // Phys. Part. Nucl. Lett. 2013. V. 10. P. 91-93].

\bibitem{Cogn}
\emph{Cognola G., Vanzo L., Zerbini S.}
Relativistic wave mechanics of spinless particles in a curved space-time //
Gen.\ Rel.\ Grav.\ 1986. V. 18. P. 971-982.

\bibitem{HN}
\emph{Hehl F.\,W. and Ni W.\,T.}
  Inertial effects of a Dirac particle //
  Phys.\ Rev.\ D. 1990. V. 42. P. 2045-2048.
  %%CITATION = PHRVA,D42,2045;%%

\bibitem{Noninertial}
\emph{Misner C.\,W.,
Thorne K.\,S., Wheeler J.\,A.} Gravitation. San
Francisco: Freeman, 1973. P. 175; \emph{Goldstein H., Poole C.\,P., Safko J.\,L.}
Classical Mechanics. 3rd edition. San
Francisco: Addison-Wesley, 2001. P. 175.

\bibitem{Tagirov}
\emph{Il'in S.B. and Tagirov E.A.} Equation of motion of a point source of a
scalar field in the general theory of relativity // Teor. Mat.
Fiz. 1978. V. 37. P. 74-83 [Theor. Math. Phys. 1978. V. 37. P. 885-891];
\emph{Tagirov E.A.}
Quantum mechanics in Riemannian spacetime.
I. Generally covariant Schr\"{o}dinger equation with
relativistic corrections // Teor. Mat.
Fiz. 1990. V. 84. P. 419-430 [Theor. Math. Phys. 1990. V. 84. P. 966-974];
%, No. 3, pp.  September, 1990.
\emph{Tagirov E.A.} Quantum Mechanics in Riemannian Space–Times.
I. The Canonical Approach //  Grav. Cosmol. 1999. V. 5. P. 23-30; \emph{Tagirov E.A.}
Quantum Mechanics in Riemannian Space: Different Approaches to Quantization of the Geodesic Motion Compared // Teor. Mat. Fiz. 2003. V. 136. P. 209-230
[Theor. Math. Phys. 2003. V. 136. P. 1077-1095].
\end{thebibliography}
\end{document}